\documentstyle[preprint,prl,aps]{revtex}
\tighten
\begin{document}
\draft
\preprint{IFP-UNC-527}
\title{Conformal ``thin sandwich'' data for the 
     initial-value problem of general 
     relativity\footnote{\bf Dedicated to John Archibald Wheeler
     and to the memory of Andr\'{e} Lichnerowicz.}}
\author{James W. York, Jr.\cite{address}}
\address{Department of Physics, North Carolina State University, 
                Raleigh, NC 27695-8202}
\date{October 15, 1998}
\maketitle

\begin{abstract}
The initial-value problem is posed by giving a conformal three-metric
on each of two nearby spacelike hypersurfaces, their proper-time separation
up to a multiplier to be determined, and the mean (extrinsic)
curvature of one slice. The resulting equations have the {\it same}
elliptic form as does the one-hypersurface formulation.
The metrical roots of this form are revealed by a 
conformal ``thin sandwich'' viewpoint coupled with the
transformation properties of the lapse function.
\end{abstract}

In this paper I propose a new interpretation of the four 
Einstein vacuum initial-value constraints. (The presence of matter would
add nothing new to the analysis.)  Partly in the spirit of a 
``thin sandwich'' viewpoint, I base this approach on prescribing the {\it conformal}
metric~\cite{York72} on each of two nearby spacelike hypersurfaces (``time slices''
$t=t^\prime \mbox{ and } t=t^\prime + \delta t$) 
that make a ``thin sandwich'' (TS). Essential use is made  of a new 
understanding of the role of 
the lapse function in general relativity~\cite{AAJWY98,YorkFest}. The new formulation 
could prove useful both conceptually, and in practice, as a way to construct
initial data in which one has a hold on the input data different from that in the currently 
accepted approach. The new approach allows us to {\it derive} from 
its dynamical and metrical foundations
the important scaling law $\bar{A}^{i j} = \psi^{-10} A^{i j}$
for the traceless
part of the extrinsic curvature. This rule is simply postulated in the
one-hypersurface approach.

The new formulation differs from the well-known TS conjecture of 
Baierlein, Sharp, and Wheeler (BSW), in which the 
{\it full} spatial Riemannian metric 
$\bar{g}_{i j}$ is given on each of two infinitesimally separated 
hypersurfaces~\cite{BSW,Wheeler,MTW}. 
(The orthogonal separation $\bar{N} \delta t$ 
between the slices is assumed never to change signs
in the BSW proposal and also here.)  
The four unknowns needed to solve the constraints were taken by BSW to be the 
``lapse function'' $\bar{N}(x)$ and the spatial ``shift vector'' 
$\bar{\beta}^i(x)$ (see below).
By using a known vacuum spacetime solution of Einstein's equations from 
which to obtain BSW data, one sees that their proposal must sometimes work.  
However, an analysis of the BSW proposal by Bartnik and Fodor~\cite{Bartnik} 
describes the general situation clearly, and one can only conclude that 
the BSW proposal is unsatisfactory.  For example, an infinite 
number of non-trivial counterexamples 
to the BSW conjecture, based on compact three-geometries of negative
scalar curvature with one 
(not $\infty^1$) constraint (fixed volume), have been 
described in~\cite{York83}.  

The initial-value problem (IVP), that is, satisfying the four constraints, 
is fundamentally a {\it one}-hypersurface embedding problem.  The four  
constraints are the Gauss-Codazzi embedding equations for a time slice 
in a Ricci-flat spacetime.  They limit the allowed values of the metric 
$\bar{g}_{i j}$ and extrinsic curvature $\bar{K}_{i j}$ of an ``initial'' time slice in a 
yet-to-be constructed vacuum spacetime.  This basic form will be referred 
to as the $(\Sigma, \bar{\mbox{\bf g}}, \bar{\mbox{\bf K}})$ form, where 
$\Sigma$ is the slice, say $t=t^\prime$.  
In this case, the constraints have already been posed as a 
semi-linear elliptic system for spatial scalar and spatial vector potentials, 
generalizations of the Newtonian potential~\cite{CBYHeld,OMY,York79}.  
A significant virtue of the formulation in this paper is that the constraints again 
become a semi-linear elliptic system with the {\it same} essential mathematical 
structure as has the $(\Sigma, \bar{\mbox{\bf g}}, \bar{\mbox{\bf K}})$ 
form.  This surprising result, as we shall see, arises from the behavior of the 
lapse function~\cite{AAJWY98,YorkFest}.

The constraint equations on $\Sigma$ are, in vacuum, 
\begin{eqnarray}
\bar{\nabla}_j(\bar{K}^{i j}-\bar{K}\bar{g}^{i j})&=&0 \; , \label{Eq:MomCon}\\
R(\bar{g})-\bar{K}_{i j}\bar{K}^{i j}+\bar{K}^2&=&0\;, \label{Eq:HamCon}
\end{eqnarray}
where $R(\bar{g})$ is the scalar curvature of $\bar{g}_{i j}$, 
$\bar{\nabla}_j$ is the Levi-Civita connection of $\bar{g}_{i j}$; 
and $\bar{K}$ is the trace of $\bar{K}_{i j}$, also called the 
``mean curvature'' of the slice.  (A review of this geometry is given 
in~\cite{York79}.)  The overbar is used to denote quantities that satisfy 
the constraints.  

The time derivative of the spatial metric $\bar{g}_{i j}$ is related to 
$\bar{K}_{i j}$, $\bar{N}$, and the shift vector $\bar{\beta}^{i}$ by  
\begin{equation}
\partial_t \bar{g}_{i j} \equiv \dot{\bar{g}}_{i j} 
	= -2\bar{N} \bar{K}_{i j} + 
(\bar{\nabla}_i \bar{\beta}_{j}+\bar{\nabla}_j \bar{\beta}_{i}) \; ,
\label{Eq:gdot}
\end{equation}
where $\bar{\beta}_{j}=\bar{g}_{j i} \bar{\beta}^{i}$.  
The fixed spatial coordinates $\vec{x}$ of a
point on the ``second'' hypersurface,
as evaluated on the ``first'' hypersurface, are displaced by 
$\bar{\beta}^i (\vec{x}) \delta t$ with respect to those 
on the first hypersurface, with an orthogonal link from the 
first to the second surface as a fiducial reference:  
$\bar{\beta}_{i}= \mbox{\boldmath $\frac{\partial}{\partial t} $} * 
\mbox{\boldmath $\frac{\partial}{\partial x^i} $}$, 
where $*$ is the physical spacetime inner product of the 
indicated natural basis four-vectors.  The essentially arbitrary 
direction of {\boldmath $\frac{\partial}{\partial t}$} is why $\bar{N}(x)$ 
and $\bar{\beta}^{i}(x)$ appear in the TS formulation.  In contrast, the tensor 
$\bar{K}_{i j}$ is always determined by the behavior of the 
unit normal on one 
slice and therefore does not possess the kinematical freedom, {\it i.e.} 
the gauge variance, of {\boldmath $\frac{\partial}{\partial t}$}. Therefore,
$\bar{N}$ and $\bar{\beta}^i$ do not appear in the one-hypersurface 
IVP for $(\Sigma, \bar{\mbox{\bf g}}, \bar{\mbox{\bf K}})$.

Turning now to the conformal metrics in the IVP, 
we recall that two metrics $g_{i j}$ and $\bar{g}_{i j}$ are conformally 
equivalent if and only if there is a scalar $\psi > 0$ such that 
$\bar{g}_{i j} = \psi^4 g_{i j}$.  The  conformally invariant 
representative of the entire conformal equivalence class, in three 
dimensions, is the weight $(-2/3)$ unit-determinant ``conformal metric'' 
$\hat{g}_{i j}=\bar{g}^{-1/3} \bar{g}_{i j}=g^{-1/3} g_{i j}$ with 
$\bar{g}=\det(\bar{g}_{i j})$ and $g=(\det g_{i j})$.  
Note particularly that for any small perturbation, 
$\bar{g}^{i j} \delta \hat{g}_{i j}=0$.  We will use the important relation 
\begin{equation}
\bar{g}^{i j} \partial_t \hat{g}_{i j} = g^{i j} \partial_t \hat{g}_{i j} 
     = \hat{g}^{i j} \partial_t \hat{g}_{i j} = 0\;.
\label{Eq:ggdot}
\end{equation}

In the following, rather than use the mathematical apparatus associated 
with conformally weighted objects such as $\hat{g}_{i j}$, 
we find it simpler to use ordinary 
scalars and tensors to the same effect. Thus, let the role of 
$\hat{g}_{i j}$ on the first surface be played by a given metric
$g_{i j}$ such that the physical metric that satisfies the constraints
is $\bar{g}_{i j} = \psi^4 g_{i j}$ for some scalar $\psi > 0$. (This
corresponds to ``dressing'' the initial unimodular conformal metric
$\hat{g}_{i j}$ with the correct determinant factor 
$\bar{g}^{1/3} = \psi^4 g^{1/3}$. 
This process does not alter the conformal equivalence class of the
metric.) The role of the conformal metric on the second surface is
played by the metric $g^\prime_{i j} = g_{i j} + u_{i j} \delta t$, 
where, in keeping with~(\ref{Eq:ggdot}), the velocity tensor
$u_{i j}= \dot{g}_{i j}$ is chosen such that 
\begin{equation}
g^{i j} u_{i j} = g^{i j} \dot{g}_{i j} = 0 \; .
\end{equation}
Then, to first order in $\delta t$, $g^\prime_{i j}$ and $g_{i j}$
have equal determinants, as desired; but $g_{i j}$ and $g^\prime_{i j}$
are not in the same conformal equivalence class in general.

We now examine the relation between the covariant derivative operators
$\nabla_i$ of $g_{i j}$ and $\bar{\nabla}_i$ of $\bar{g}_{i j}$.
The relation is determined by
\begin{equation}
\bar{\Gamma}^i\mathstrut_{j k}(\bar{g}) = \Gamma^i\mathstrut_{j k}(g) 
	+ 2 \psi^{-1} \left( 2 \delta^i_{( j} \partial_{k )} \psi 
	- g^{i l} g_{j k} \partial_l \psi \right) \; ,
\end{equation}
from which follows the scalar curvature relation first used in an
initial-value problem by Lichnerowicz~\cite{Lich},
\begin{equation}
R(\bar{g}) = \psi^{-4} R(g) - 8 \psi^{-5} \Delta_g \psi \; ,
\end{equation}
where $\Delta_g \psi \equiv g^{k l} \nabla_k \nabla_l \psi$ is
the ``rough'' scalar Laplacian associated with $g_{i j}$.

Next, we solve~(\ref{Eq:gdot}) for its traceless part 
\begin{equation}
\dot{\bar{g}}_{i j} - \frac{1}{3} \bar{g}_{i j} \bar{g}^{k l} 
	\dot{\bar{g}}_{k l} \equiv \bar{u}_{i j} 
	= -2 \bar{N} \bar{A}_{i j} + (\bar{L} \bar{\beta})_{i j}
\label{Eq:traceless}
\end{equation}
with $\bar{A}_{i j} \equiv \bar{K}_{i j} - \frac{1}{3} \bar{K} \bar{g}_{i j}$
and 
\begin{equation}
(\bar{L} \bar{\beta})_{i j} \equiv \bar{\nabla}_i \bar{\beta}_j
	+ \bar{\nabla}_j \bar{\beta}_i - (2/3) \bar{g}_{i j}
	\bar{\nabla}^k \bar{\beta}_k \; .
\label{Eq:LB}
\end{equation}

Expression~(\ref{Eq:LB}) vanishes, for non-vanishing $\bar{\beta}^i$, if
and only if $\bar{g}_{i j}$ admits a conformal Killing vector
$\bar{\beta}^i = k^i$. Clearly, $k^i$ would also be a
conformal Killing vector of $g_{i j}$,
or of any metric conformally equivalent to $\bar{g}_{i j}$, with no
scaling of $k^i$. This teaches us that in general $\bar{\beta}^i = \beta^i$,
while $\bar{\beta}_i = \bar{g}_{i j} \bar{\beta}^j = \psi^4 g_{i j} \beta^j=\psi^4 \beta_i$.
That $\bar{\beta}^i=\beta^i$ also follows because $\beta^i$, generator
of a spatial diffeomorphism, is not a dynamical variable. The latter
``rule'' was inferred as a matter of principle.

It is clear in~(\ref{Eq:traceless}) that the left hand side $\bar{u}_{i j}$ satisfies
$\bar{u}_{i j} = \psi^4 u_{i j}$ because the terms in $\dot{\psi}$ cancel
out. Furthermore, a straightforward calculation shows that
\begin{equation}
(\bar{L} \bar{\beta})_{i j} = \psi^4 (L \beta)_{i j} \; ; \qquad
(\bar{L} \beta)^{i j} = \psi^{-4} (L \beta)^{i j} \; .
\end{equation}

Next, we note, perhaps surprisingly,
that the lapse function $\bar{N}$ has essential 
non-trivial conformal behavior.
Furthermore, this is {\it the} new element in the IVP analysis.
In~\cite{Teitel,Ashtekar,AAJWY98,YorkFest} the ``slicing function''
$\alpha(t,x) > 0$ replaces the lapse function $\bar{N}$,
\begin{equation}
\bar{N} = \bar{g}^{1/2} \alpha \; ,
\label{Eq:slicingfunction}
\end{equation}
with important improvements then appearing in Teitelboim's
path integral~\cite{Teitel}, in Ashtekar's new variables
program~\cite{Ashtekar}, in the canonical action 
principle~\cite{AAJWY98,YorkFest}, and in making clear the 
role of the contracted Bianchi identities~\cite{AAJWY98,YorkFest}.
The lapse is now a dynamical variable because of the
$\bar{g}^{1/2}$ factor~\cite{Ashtekar,AAJWY98,YorkFest}.
Furthermore, in the construction of
mathematically hyperbolic systems for the Einstein {\it evolution}
equations with explicitly physical characteristics, and only such,
(for example~\cite{CBY97,ACBY97,CBYAnew}), it turns out to be
$\alpha(t,x)$, not the usual lapse function $\bar{N}$, that can
be freely specified. This use of $\bar{N} = \bar{g}^{1/2} \alpha$ is
Choquet-Bruhat's ``algebraic gauge''~\cite{CBRug,CBY95} with, in general,
a ``gauge source''~\cite{Friedrich}. Actually, 
$\bar{N} = \bar{g}^{1/2} \alpha$ should be seen as 
a change of variables in which 
one specifies freely $\alpha(t,x)>0$ rather than $N$.
For these reasons, we conclude
that $\alpha$ is not a dynamical variable, $\bar{\alpha} = \alpha$.
For the lapse, we have from~(\ref{Eq:slicingfunction}),
with $N$ given and positive,
\begin{equation}
\bar{N} = \psi^6 N \; .
\end{equation}

Finally, we recall from the standard initial value problem for
$(\Sigma,\bar{\mbox{\bf g}}, \bar{\mbox{\bf K}})$, that the separation of the
extrinsic curvature into (its irreducible) trace and 
traceless parts is fundamental,
as it is here, and that $\bar{K}=K$~\cite{York72}: the trace is
not transformed even though it is dynamical. It ``anchors'' the
construction, setting a reference scale by fixing an observable
dimensionful dynamical variable.
(In closed worlds $K$ is like a ``time'' variable,
in that it may ``locate'' the thin sandwich. In cosmology, $K$ is
essentially the inverse mean ``Hubble time.'') There is no
underlying geometrical derivation of $\bar{K}=K$, unlike the case of
$\bar{A}_{i j}$ below. The conformal invariance of $K$ is primitive.
See the result in~(\ref{Eq:DotLogPsi}) below.

Now we solve~(\ref{Eq:traceless}) for $\bar{A}^{i j}$ and find
\begin{eqnarray}
\bar{A}^{i j} &=& \psi^{-6} (2N)^{-1} \left[ \psi^{-4} (L \bar{\beta})^{i j}
     - \psi^{-4} u^{i j} \right] \nonumber\\
&=& \psi^{-10} \left\{ (2N)^{-1} \left[ (L \bar{\beta})^{i j} - u^{i j} \right] \right\}
     = \psi^{-10} A^{i j} \; ,
\label{Eq:Abar}
\end{eqnarray}
the same conformal scaling that was postulated 
by Lichnerowicz~\cite{Lich} and others~\cite{CBYHeld,OMY,York79} 
for the traceless part
of $\bar{K}^{i j}$ in the one-hypersurface 
problem. One now has a derivation of this
fundamental transformation from its metrical foundations.
The momentum constraint~(\ref{Eq:MomCon}) becomes
\begin{eqnarray}
\nabla_j \left[ (2N)^{-1} (L \bar{\beta})^{i j}\right] &=& 
     \nabla_j \left[ (2N)^{-1} u^{i j} \right] \nonumber\\
     & & + (2/3) \psi^6 \nabla^i K \; ,
\label{Eq:NewMomCon}
\end{eqnarray}
for unknown $\bar{\beta}^i$ and known $N$, $g_{i j}$, 
$u_{i j}$, and $K$. The operator on the left, being in elliptic
``divergence form'' with $N>0$, does not differ in any important
property from its counterpart in the 
$(\Sigma, \bar{\mbox{\bf g}}, \bar{\mbox{\bf K}})$ 
analysis~\cite{CBYHeld,OMY,York79}. The Hamiltonian 
constraint~(\ref{Eq:HamCon}) becomes~\cite{York73}
\begin{equation}
8 \Delta_g \psi - R(g) \psi + A_{i j} A^{i j} \psi^{-7} - (2/3) K \psi^5 = 0 \; ,
\label{Eq:NewHamCon}
\end{equation}
for unknown $\psi$, where $A^{i j}$ is given in~(\ref{Eq:Abar}). This equation
has precisely the same form in the one-hypersurface and two-hypersurfaces
constraint problems. Note that~(\ref{Eq:NewMomCon}) 
and~(\ref{Eq:NewHamCon}) are not coupled if $K = \mbox{constant}$,
{\it i.e.}, one solves~(\ref{Eq:NewMomCon}), then~(\ref{Eq:NewHamCon}).
{\it No tensor splittings}~\cite{York73,York74} are needed in 
the new formulation of the constraints.

Thus, the free data are $\left\{g_{i j}, u_{i j}, N, K\right\}$ 
and the solution is $\left\{ \psi, \bar{\beta}^i \right\}$. 
Mathematical analysis of the corresponding elliptic 
system~(\ref{Eq:NewMomCon}, \ref{Eq:NewHamCon})
has been carried out elsewhere, for 
example,~\cite{CBYHeld,OMY,Isenberg,CBBrill,CBIJWY}, and 
will not be repeated here. The corresponding situation in the 
$(\Sigma,\bar{\mbox{\bf g}}, \bar{\mbox{\bf K}})$ analysis is that
the free data are  $\left\{g_{i j}, A_{i j}, K\right\}$
and the solution is $\left\{ \varphi, W^i \right\}$, where $W^i$ is
obtained from a tensor splitting of $A_{i j}$~\cite{York73,York74}.
Note that $\varphi \neq \psi$ and $W^i \neq \bar{\beta}^i$.
Only part of $A_{i j}$, found in the splitting, is free.
The conformal covariance of the new method, {\it i.e.}, starting
with different representatives of a given conformal equivalence
class is {\it unique} and clear. On the other hand, that of the
$(\Sigma,\bar{\mbox{\bf g}}, \bar{\mbox{\bf K}})$ analysis can 
follow two inequivalent routes because there are two slightly
different conformal analyses possible for construction
of $\left( \Sigma, \bar{\mbox{\bf g}}, \bar{\mbox{\bf K}} \right)$.
This non-uniqueness arises because conformal scaling and tensor
splittings are not commutative in a straightforward way.
The method of tensor splitting in~\cite{York79} gives the
Hamiltonian constraint in the form of~(\ref{Eq:NewHamCon}).

These data are not in perfect analogy to those conjectured
by BSW, because $K$ and $N$ can be thought of
as belonging to the thin sandwich as
a whole. The role of $K$ has been described. The role of 
$N = g^{1/2} \alpha$ is to give the thickness of the sandwich,
$\bar{N} \delta t$, in proper time measured orthogonally 
from $t=t^\prime$ to $t=t^{\prime\prime}$:
\begin{equation}
\bar{N} \delta t = (\bar{g}^{1/2} \alpha) \delta t 
     = (\psi^6 g^{1/2}) \alpha \delta t = \psi^6 (N \delta t) \; .
\end{equation}

The final relationships between the two 
physical Riemannian metrics 
$\bar{g}_{i j}$ and $\bar{g}^\prime_{i j} = \bar{g}_{i j} + 
\dot{\bar{g}}_{i j} \delta t$ and the given data 
$g_{i j}$ and $g^\prime_{i j} = g_{i j} + u_{i j} \delta t$
are quite interesting. Of course, $\bar{g}_{i j} = \psi^4 g_{i j}$
is clear. But we have to calculate the relationship between $\bar{g}_{i j}$ 
and $\bar{g}^\prime_{i j}$ as
$\bar{g}^\prime_{i j} = \bar{g}_{i j} + \dot{\bar{g}}_{i j} \delta t$,
where, as in~(\ref{Eq:gdot}),
\begin{eqnarray}
\dot{\bar{g}}_{i j} &=& \partial_t \left( \psi^4 g_{i j} \right) \nonumber\\
     &=& -2 \bar{N} \left( \bar{A}_{i j} + \frac{1}{3} \bar{g}_{i j} K \right)
     + \left( \bar{\nabla}_i \bar{\beta}_j + \bar{\nabla}_j 
     \bar{\beta}_i \right) \; .
\label{Eq:gbardot}
\end {eqnarray}
Working out (\ref{Eq:gbardot}) gives a key result, namely,
\begin{eqnarray}
\dot{\bar{g}}_{i j} &=& \psi^4 \left[ u_{i j} + 
     g_{i j} \partial_t \left(4 \log \psi\right) \right] \nonumber\\
     &=& \bar{u}_{i j} + \bar{g}_{i j} \partial_t \left(4 \log \psi\right) \; ,
\end{eqnarray}
where
\begin{eqnarray}
\partial_t \left( 4 \log \psi \right) &=& \frac{2}{3} \left( \nabla_k \beta^k + 
     6 \beta^k \partial_k \log \psi -  N K \psi^6 \right) \nonumber\\
&=& \partial_t \left( \bar{g}/g \right)^{1/3} = \frac{2}{3} 
     \left( \bar{\nabla}_k \bar{\beta}^k - \bar{N} \bar{K} \right) \; .
\label{Eq:DotLogPsi}
\end{eqnarray}
Therefore,
\begin{eqnarray}
\dot{\bar{g}}_{i j} &=& \psi^4 \left[ u_{i j} + \frac{2}{3} g_{i j}
	\left( \nabla_k \beta^k + 6 \beta^k \partial_k \log \psi
	- N K \psi^6 \right) \right] \nonumber\\
&=& \bar{u}_{i j} + \frac{1}{3} \bar{g}_{i j} 
	\left( 2 \bar{\nabla}_k \bar{\beta}^k - 2 \bar{N} K \right) \; .
\label{Eq:Bargdot}
\end{eqnarray}
We see that $\dot{\psi}$ and $\dot{\bar{g}}_{i j}$ are 
fully determined by the constraints and, in the last equality
of~(\ref{Eq:Bargdot}), that
the conformal invariance of $\beta^k$ ($=\bar{\beta}^k$)
and $K$ ($= \bar{K}$) are fully
consistent, having led to the precisely geometrically
correct form of $\dot{\bar{g}}_{i j}$ by virtue 
also of $\bar{N} = \psi^6 N$.

This interpretation of the semi-linear elliptic constraint system has
interesting differences from earlier ones because the data and solutions
are related more simply to the spacetime metric, though
not in the manner that would be implied by ordinary conformal 
transformations of the spacetime metric.
In this ``conformal'' TS form one can see explicitly the role of
every part of the metric. The new
formulation shows that the one-hypersurface and
two-hypersurfaces initial-value problems are both viable
once the full implications in general relativity
of the ``dynamical conformal structures'' are understood. The two
viewpoints can be thought of as roughly analogous to a 
Hamiltonian and to a Lagrangian view of the constraints; 
the former because using $\bar{K}_{i j}$
directly~\cite{CBYHeld,OMY,York79} is equivalent to using 
the initial canonical momentum $\bar{\pi}^{i j} = \bar{g}^{1/2}
\left( \bar{K} \bar{g}^{i j} - \bar{K}^{i j} \right)$, and the
latter because $\dot{\bar{g}}_{i j}$ is the initial velocity. 
This striking correspondence hangs on the subtle role of the lapse
function through the Choquet-Bruhat relation
$\bar{N} = \bar{g}^{1/2} \alpha$ and on the corresponding
conformal invariance of $K$ postulated by the author~\cite{York73}
in going beyond Lichnerowicz's choice $K=0$. The 
``conformal thin sandwich'' aspect of the results reflects
Wheeler's approach.

\begin{acknowledgements}
I am grateful to A. Anderson, J. David Brown, 
N. O'Murchadha, and especially
Y. Choquet-Bruhat for encouragement and to Sarah and Mark
Rupright for technical assistance. I owe special 
thanks to Dean Jerry Whitten of the College
of Physical and Mathematical Sciences of North 
Carolina State University for making my Leave of Absence possible,
and to J.~A.~Isenberg for his advice on the presentation. 
Research support has been received by the author from National Science 
Foundation Grants No.~PHY~94-13207 to the University of North Carolina,
Chapel Hill,
and No.~PHY~93-18152/ASC~93-18152 (ARPA supplemented).
\end{acknowledgements}

\end{document}